\documentclass[%
 reprint,
 amsmath,amssymb,
 aps,
]{revtex4-2}

\usepackage[usenames]{color}
\usepackage{colortbl}
\usepackage{ulem}

\usepackage{graphicx}% Include figure files
\usepackage{dcolumn}% Align table columns on decimal point
\usepackage{bm}% bold math

\begin{document}

\preprint{APS/123-QED}

\title{Nonlinear screening and charge redistribution in periodically doped graphene}% Force line breaks with \\

\author{K.A. Baryshnikov}
\author{A.V. Gert}%
 \email{anton.gert@mail.ioffe.ru}
\author{Yu.B. Vasilyev}%
\author{A.P. Dmitriev}%
\affiliation{Ioffe Institute, 194021 St. Petersburg, Russia}

\date{\today}% It is always \today, today,

\begin{abstract}
The screening problem for the Coulomb potential of a charge located in a two-dimensional (2D) system has an intriguing solution with a power law distance screening factor due to out-of-plane electrical fields. This is crucially different from a three-dimensional case with exponential screening. 
The long-range action of electric fields results in the effective inflow of electrons from high-doped regions to low-doped regions of a 2D heterostructure.
In graphene and other materials with linear energy spectrum for electrons, such inflow in low-doped regions also occurs, but its effectiveness is dependent on doping level. 
This can be used for fabricating high-mobility conducting channels.
We provide the theory for determining electron potential and concentration in a periodically doped graphene sheet along one dimension taking into account all effects of long-range 2D screening. 
This results in a substantially nonlinear integro-differential problem, which is solved numerically via computationally cheap algorithm. Similar nonlinear problems arise in a wide range of doped 2D heterostructures made of linear spectrum materials. 

\end{abstract}

\maketitle

%\tableofcontents
\section{\label{sec:level1}Introduction}

The possibilities of using 2D superlattices are based on the large surface-to-volume ratio, which is important for creating batteries, capacitors, emitters and detectors of electromagnetic waves, as well as devices based on current instabilities and plasmonic effects \cite{A0}. This is also true for graphene, a promising 2D material for nano-optics \cite{A1,A2,A3,A4}. Periodically doped graphene sheets are of interest for creating and optimizing metasurfaces that allow manipulation of radiation in the terahertz range using a single atomic layer \cite{A5,A6,A7,A8,A9,A10,A11}. 
Additionally, for one-dimensional periodic potentials in graphene, the theory predicts conductivity oscillations corresponding to the appearance of additional Dirac points and Van Hove singularities in the density of states \cite{Park,Abed,Brey,Barb}, lensing \cite{B3}, supercollimation \cite{B4,B5} of electrons,  and Klein tunneling effects \cite{B10,B11}. Therefore, reports on the experimental implementation of graphene superlattices have attracted much attention \cite{Meyer,Vazq,Ma,Tan,Wang,Barb1}.

The simplest example of a lateral graphene superlattice is a structure with periodic modulation of the density of states along the graphene sheet by a series of metal gates \cite{Dubey,Drien,Olbr}. Structured dielectric matrices \cite{Li,Fors,Giul}, ferroelectric gates with periodically structured domains \cite{Tian}, and periodic modulation of the electron gas density caused by the illumination of special films on the graphene surface with ultraviolet radiation \cite{Lara} can also be used. To form superlattices, selective ultraviolet irradiation was also used to induce p-type doping in an intrinsic n-type epitaxial graphene monolayer grown on a SiC substrate \cite{vasilyev2018high}. The latter method is equivalent to the efficient lateral periodic doping of a graphene sheet (without any metal gates or external electric fields) to form a series of p-n junctions along it, i.e. alternating p- and n-type regions. The same method can also be used to form alternating high and low doped regions of the same type (see below).

In theoretical studies of superlattices in 2D structures \cite{B4,B5,B10,B11,B12,B13}, the periodic potential is typically assumed to be strictly step-shaped, although this is not always an accurate approximation. The concentration of charge carriers in the superlattice varies greatly, causing flow from regions with high concentration to regions with low concentration. This flow is limited by the screening effect, which differs from the exponential behavior seen in 3D cases and instead has a power-law dependence on distance \cite{keldysh2024coulomb,rytova,stern1967polarizability} (known as the Rytova-Keldysh potential). As a result, the carrier concentration and the periodic potential of the superlattice change very smoothly, particularly in structures with narrow regions of low carrier concentration.

The impact of 2D screening on the potentials of static charges in graphene and on the effective dielectric function of carriers was theoretically studied in \cite{C0,C1,C2,C3}. Specifically, it has been demonstrated that the Thomas-Fermi screening of a sufficiently large charge in graphene is significantly nonlinear. The nonlinear screening in the context of ballistic transport through a single p-n junction was explored in a theoretical research by Zhang and Fogler \cite{C4}. The authors numerically solved the self-consistent problem of determining the carrier concentration and the electric field in the region of the single p-n junction. They obtained analytical results near the center of the p-n junction and calculated a power-law attenuation of the volume charge density far from the p-n junction. A similar issue of determining the band bending when a single metal gate is brought close to a graphene sheet was addressed in \cite{C5}.  
The researchers conducted calculations of the carrier concentration distribution and potential near the gate edge, and also derived the asymptotic law for these quantities far from the gate edge. However, the same effects of charge redistribution and self-consistent potential determination in graphene heterostructures and superlattices are still being investigated. 

The litarature discusses the effects associated with the power-law nature of screening in 2D structures for conventional materials with a quadratic spectrum of electron kinetic energies. For example, these effects analyzed for potential profile calculations of a 2D electron gas in GaAs/Al$_x$Ga$_{1-x}$As quantum-well-based gated lateral superlattices \cite{DaviesLarkin1994theory}.
However, more illustrative results can be found in a study by Dmitriev and Shur \cite{dmitriev2013lateral} for an $n_+$–$n_0$–$n_+$ heterostructure made of conventional semiconductor materials with a quadratic energy spectrum.
The study showed that the electron concentration in the low-doped $n_0$ region decreases from the edges to the center following a power law with a characteristic distance scale equal to the Bohr radius $a_B = \varepsilon \hbar^2/m e^2$. This scale depends solely on the material parameters of the structure and fundamental constants. 
The electron concentration in the $n_0$ region increases as the dielectric constant $\varepsilon$ of the surrounding material increases, leading to a redistribution of electrons from the $n_+$ region. If a metal gate is placed parallel and close to the structure (equivalent to setting $\varepsilon = \infty$), the electron concentration in the $n_0$ region becomes equal to that in the $n_+$ region, forming a channel with high conductivity.

In this work, we present a theoretical study of the Coulomb potential and electron density redistribution in a 1D periodically doped graphene. 
In contrast to \cite{dmitriev2013lateral}, our theory considers all the effects of long-range 2D screening for periodic heterostructures and the influence of the linear spectrum, yielding an essentially nonlinear problem that can only be solved analytically in specific cases.
We have developed a computationally efficient algorithm for the self-consistent determination of the electronic potential and electron concentration in periodically doped graphene, and we have provided the results of a numerical solution to the problem. Our findings demonstrate that the effective shielding length is dependent on the doping level, consistent with previous results for Rytova-Keldysh shielding of point charges in graphene \cite{C1,C2}. 
Additionally, we have shown that screening effects for a periodic superlattice (periodic doping of one type) in graphene result in a significant inflow of electrons in regions of low doping, which can be used for the formation of high mobility channels on a chip without the need for gates. 
Finally, we emphasize that the nonlinear nature of screening effects leads to a renormalization of the average potential, influenced by all Fourier components of the potential that are self-consistently determined with the distribution of electrons in the sample. This is a crucial consideration for the design of superlattices based on graphene or other materials with a linear spectrum.

\section{\label{sec:level1}Theory}
\begin{figure}[htp]
\includegraphics[scale=0.23]{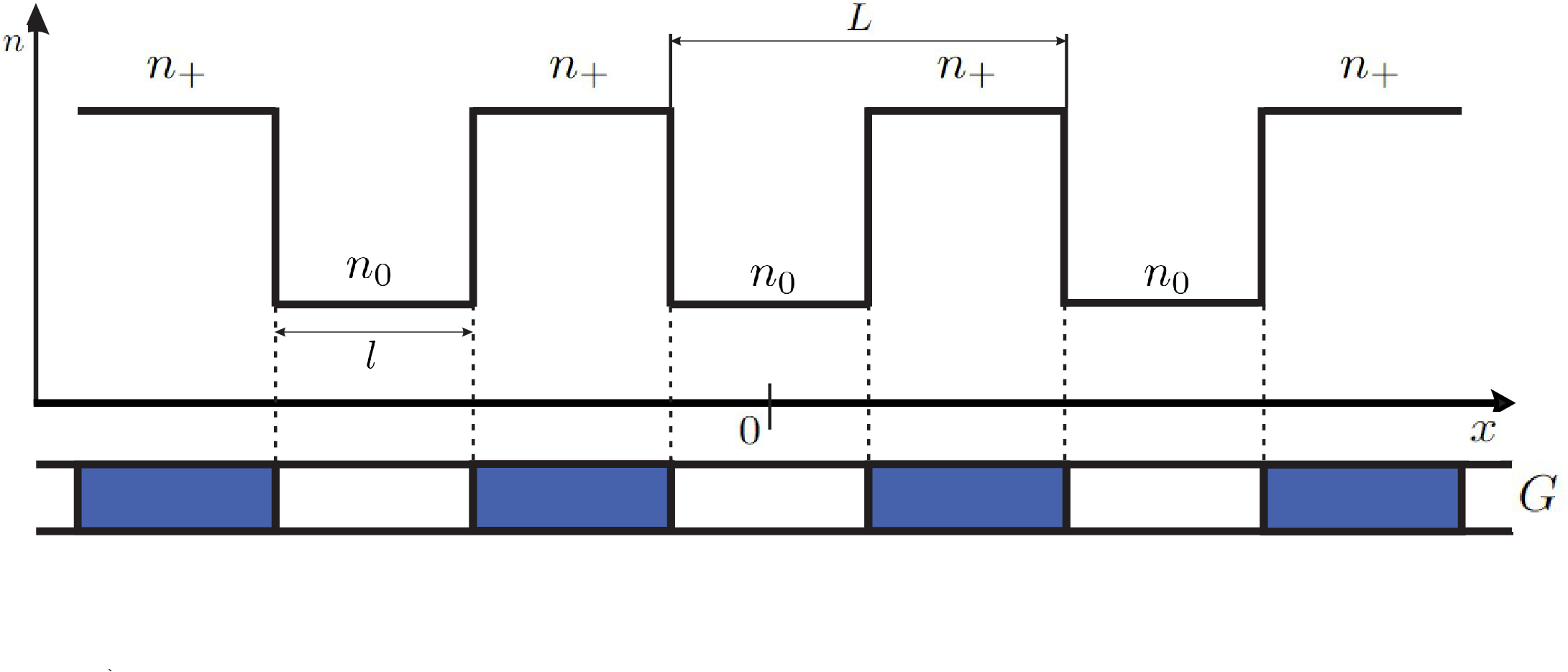}
\caption{\label{GSL} An illustration of a graphene sheet (side view) that is periodically doped with different concentrations of donors along the $x$ axis. The doping period is denoted by $L$. Regions with high doping levels ($n_+$) are colored blue, while regions with low doping levels ($n_0$, having the length of $l$) are colored white. }
\end{figure}

\subsection{\label{sec:level2} Periodically doped graphene}

graphene sheet laterally doped by donors along the $x$-dimension with a period of $L$ is shown in Fig.~\ref{GSL}. Each doping region is a stripe of infinite width. Low-doped regions (with donor concentration $n_0$) have a length of $l$ and high-doped regions (with donor concentration $n_+$) have a length of $L-l$.
We are interested in the equilibrium distributions of electron density in such a structure, determined by the requirement that the Fermi level is identical at all points of the structure. The Fermi level is established as a result of the flow of electrons from more doped regions to less doped regions. In this case, regions with a charge density different from zero are formed. 

Let us assume that the 2D sample sheet is located within a dielectric with a constant $\varepsilon$. The field potential at a point is determined by the solution of Poisson equation $\varphi({\bm r}) = - e \int d{\bm r'} (n({\bm r'})-N_{\rm D}({\bm r'}))/\varepsilon |{\bm r} - {\bm r'}| $, where $e$ represents the absolute value of the electron charge, $N_{\rm D}$ is the donor concentration that sharply changes at the boundaries of high and low doped regions, and $n({\bm r'})$ is the electron concentration distribution.

Assuming an infinitely small thickness of the graphene sheet along the $z$ direction (its value is assumed to be the smallest length parameter in all calculations) and implying homogeneity in the $y$ direction (neglecting edge effects and considering the potential far away from the edges of the graphene sheet), one can reduce the problem of determining potential changes along the $x$ direction of the graphene sheet to a 1D integro-differential equation \cite{dmitriev2013lateral}. This reduction results in a logarithmic kernel, which is sufficient for a finite $n_+$--$n_0$--$n_+$ heterostructure, but it is inappropriate for calculating an infinitely long periodic superlattice along the $x$ axis due to integral divergence. Therefore, we use the same reduction, but for the derivative of the potential, i.e., for electric fields, which results in the following integro-differential relation
\begin{equation}
\label{dphix}
    \varphi'_x (x) = \frac{2 e}{\varepsilon} {\rm v.p.} \int\limits_{-\infty}^{+\infty} \frac{n(x')-N_{\rm D}(x')}{x-x'} {\rm d}x' .
\end{equation}
This result can be considered as the sum of the electric fields at point $x$ of infinite charged rods located at points $x'$ and placed along the $y$ axis. Each rod has a width of $dx'$ and a linear charge density of $\tau = -e (n(x')-N_{\rm D}(x'))dx'$. By surrounding each rod with a cylinder of radius $r$ and using Gauss theorem, we get the electric field ${\bm E} = 2\tau {\bm r}/\varepsilon r^2$ at a radius vector ${\bm r}$ from this individual rod in the medium with dielectric constant $\varepsilon$. After making obvious substitutions $E_{x} = -d\varphi/dx$ and $r_{x}/ r^2 = 1/(x-x')$ and summing up the contributions from all rods (equivalent to integration along $dx'$), we arrive at Eq.~(\ref{dphix}). The integral should be considered in a principal value sense.

To obtain the system of equations for $n(x)$ and $\varphi(x)$, we need to establish an additional relationship between these quantities. In contrast to \cite{dmitriev2013lateral}, we consider the linear spectrum of kinetic energy of electrons, given by $E(x,p)=vp-e\varphi(x)$, where $v=10^8$~cm/s is the electron velocity in graphene. Therefore, the electron concentration reads as
\begin{equation}
 \label{nx}
     \displaystyle n (x) = 4 \int\limits_{0}^{+\infty} \!\! \int\limits_{0}^{~2\pi} \frac{p {\rm d} p {\rm d} \theta}{(2\pi\hbar)^2} \frac{1}{1+\exp \left(E(x,p) - E_F \right)/T } .
 \end{equation}
Setting $T=0$, we get an asymptotic expression \cite{rhodes1950fermi,kalitkin2018fermi} 
%for Eq.~(\ref{nx})
 \begin{equation}
 \label{nxT0}
        n (x) = \gamma (e\varphi(x)+E_F)^2,
\end{equation}
that is valid when $e\varphi(x)+E_F > 0$. When $e\varphi(x)+E_F < 0$ the expression should be zero at $T=0$. Introducing a parameter $\gamma = 1/\pi v^2 \hbar^2$, which represents the square of the density of states. Hence, to determine the potential in a graphene superlattice, one must solve a self-consistent integro-differential equation for $\varphi (x)$, which is highly nonlinear.

\subsection{Electrical potential equation analysis}
It will be convenient for us to rewrite the equation for electrical potential as the potential energy change along the graphene sheet $e\varphi(x)$, plus the chemical potential $E_F$ (Fermi energy at $T=0$) of a charge in the sheet
\begin{equation}
\label{phigeneralized}
    \phi (x) = e\varphi (x) + E_F , \quad \phi'_x (x)=e\varphi' (x) .
\end{equation}
The equation for $\phi(x)$ derived from Eqs.~(\ref{dphix}) and (\ref{nxT0}) is
\begin{eqnarray}
\label{phigen-eq}
    \phi'_x (x) = \frac{2 e^2}{\varepsilon} {\rm v.p.} \int\limits_{-\infty}^{+\infty} \frac{\gamma [\phi(x')]^2-N_{\rm D}(x')}{x-x'} {\rm d}x' .
\end{eqnarray}
Due to translational symmetry along the superlattice periods we decompose all functions of $x$ into Fourier series
\begin{equation}
\label{phiFourier}
    \phi (x) = \sum\limits_k \hat{\phi}_k e^{ikx}, \quad \hat{\phi}_k = \frac{1}{L} \int\limits_{-L/2}^{L/2} \phi (x) e^{-ikx} {\rm d}x.
\end{equation}
Here, the inverse space is discrete with $k=2\pi m/L$, where $m \in {\mathbb Z}$. By definition, we imply that $\hat{\phi}_0 = E_F$ and $\hat{\phi}_k = e \hat{\varphi}_k$, while $\hat{\varphi}_0 = 0$. It is easy to show that the Fourier image of a squared function is equal to the convolution of the Fourier images of the function itself
\begin{equation}
\label{phi2Fourier}
    \left[\phi (x)\right]^2 = \sum\limits_k e^{ikx} \sum\limits_q \hat{\phi}_q \hat{\phi}_{k-q}.
\end{equation}
Therefore, Eq.~(\ref{phigen-eq}) in Fourier components reads as
\begin{equation}
\label{phiF-nonlin-appr}
    k \hat{\phi}_k = \frac{2\pi e^2}{\varepsilon} {\rm sign}(k) \left( \hat{N}_{\rm D}^{(k)} - \gamma \sum\limits_q \hat{\phi}_q \hat{\phi}_{k-q}  \right), 
\end{equation}
where we used the continuous Fourier transform image of the kernel ${\cal F}\left[\frac{1}{x}\right](k) = (-i\pi){\rm sign}(k)$, which occurs when integrating from negative infinity to positive infinity along the $x'$ axis in Eq.~(\ref{phigen-eq}).

Note that formally we cannot consider $k=0$ in Eq.~(\ref{phiF-nonlin-appr}) because it results in true equivalence of zeros on both sides of the equation due to properties of the kernel. This restriction arises from our consideration of the potential derivative (electrical field) to ensure converging integrals on the right-hand side. However, it is worth noting that ${\rm sign}(k) = k/|k|$, which, after formal division by the kernel on both sides, suggests that the expression in the large brackets in Eq.~(\ref{phiF-nonlin-appr}) should also tend to zero at $k=0$. This is true when considering the additional requirement of electroneutrality. In the case of an infinite plane, it should not be charged on average to allow for a reasonable decay of potential at infinitely large distances from the plane. Therefore, we must require that the concentration of electrons along one period equals the concentration of donors (assuming one donor gives one free electron in the system, as in our model), i.e.
\begin{equation}
\label{Parseval}
    \frac{1}{L} \int\limits_{-L/2}^{L/2} N_{\rm D}(x) {\rm d}x = \frac{1}{L} \int\limits_{-L/2}^{L/2} n(x) {\rm d}x .
\end{equation}
If we use now Eq.~(\ref{nxT0}), it results in
\begin{equation}
\label{Fourier-nD0}
    \hat{N}_{\rm D}^{(0)} =  \gamma \sum\limits_q \hat{\phi}_q \hat{\phi}_{-q} = \gamma \sum\limits_q \left[\hat{\phi}_q\right]^2 .
\end{equation}
Here we also used the Parseval-Rayleigh identity for Fourier images and the $x$-inversion symmetry to ensure that $\hat{\phi}_k = \hat{\phi}_{-k}$. As a result, we obtain a complete infinite system of nonlinear equations on Fourier images $\hat{\phi}_k$ for all $k\neq 0$ in Eq.~(\ref{phiF-nonlin-appr}), which along with Eq.~(\ref{Fourier-nD0}) defines $\hat{\phi}_0$.

Eq.~(\ref{phigen-eq}) can be linearized if we assume $|e\varphi(x)|\ll E_F$, which is equivalent to the condition $|\hat{\phi}_k| \ll |\hat{\phi}_0|$. This means that we can leave only terms proportional to $\hat{\phi}_0 \hat{\phi}_k$ in Eq.~(\ref{phiF-nonlin-appr}) and the only term $\hat{\phi}_0^2$ in Eq.~(\ref{Fourier-nD0}). One can easily check that the solution of such a linearized equation can be obtained analytically 
\begin{eqnarray}
\label{phiF-in-linappr}
    \displaystyle \hat{\phi}_k  
    =\frac{2 \pi e^2}{\varepsilon} \frac{ \hat{N}_{\rm D}^{(k)}}{|k| + l_s^{-1}}. 
\end{eqnarray}
Here, $k$ takes all nonzero discrete values and an effective screening length is introduced as $l_s = \varepsilon v^2 \hbar^2/(4 E_F e^2)$ depending on the Fermi energy $E_F = \hat{\phi}_0 = v\hbar \sqrt{\pi \hat{N}_{\rm D}^0}$, defined from Eq.~(\ref{Fourier-nD0}), where terms $\propto \hat{\phi}_{k\neq 0}^2$ are assumed to be negligibly small. Therefore, we have confirmed the dependence of the effective screening length in graphene on the doping level \cite{C1,C2}.

We assume that the doping level could be approximated by a sharp square wave structure on the scale of $100$~nm as shown in Fig.~\ref{GSL}. Then its Fourier components are
\begin{equation}
\label{nDF}
    \hat{N}_{\rm D}^{(k)} = - \frac{2\Delta N}{kL}\sin{\frac{kl}{2}} + n_{+}\delta_{k,0},
\end{equation}
where $\Delta N = n_{+} - n_0$ is the difference in doping levels between distinct regions and $\delta_{k,0}$ is the Kronecker delta symbol. It is now easy to see that the average doping level along the period is $\hat{N}_{\rm D}^0 = n_{+} - \Delta N (l/L)$, and the linear approximation solutions for $k\neq 0$ are proportional to $\Delta N$. Thus, the condition for the linear approximation $|e\varphi(x)|\ll E_F$ or $|\hat{\phi}_k| \ll |\hat{\phi}_0|$ can be rewritten in terms of doping level differences as $\Delta N \ll \hat{N}_{\rm D}^0$, meaning a weak spatial doping contrast is required. In the case of equal sizes of high-doped and low-doped regions $l=L/2$, the requirement for the linear approximation to hold is $\Delta N \ll (n_0 + n_+)/2$.

\subsection{Numerical solution by iteration method}
If one is seeking a solution for the case, where the difference in doping levels is significant and the condition $\Delta N \ll (n_{+} + n_{0})/2$ is violated, the complete set of nonlinear equations must be solved numerically. We previously introduced the inverse screening length $l_s^{-1}$, which has the same dimension as the variable $k$. Additionally, we will move to dimensionless parameters $D_k=\hat{N}_{\rm D}^k/\hat{N}_{\rm D}^0$ and variables $\psi_k = \phi_k / E_F$. 

If we start with a homogeneous graphene sheet with some a nonzero Fermi level, we set $\psi_0^{(s)} = 1$ and $\psi_{k\neq 0}^{(s)} = 0$ at the first iteration step $s=1$. In the next step, we take into account all effects of periodically doped graphene sheet self-consistently. We apply a change of variables to dimensionless ones in Eq.~(\ref{phiF-nonlin-appr}) and Eq.~(\ref{Fourier-nD0}). Hence, at the next iteration steps $s > 1$, we have 
\begin{equation}
\label{psi_k}
    \displaystyle \psi_k^{(s)} = \frac{1/2}{|k|l_s+ \psi_0^{(s)}} \left[ D_k - \sum\limits_{q\neq 0,k} \psi_q^{(s-1)} \psi_{k-q}^{(s-1)} \right]
\end{equation}
for all $k\neq 0$. Numerically, we obtain the first thousand discrete values of $k=2\pi m/L$, where $m$ ranges from $-5000$ to $5000$ at each step. One can see that $\psi_0$ is picked out and moved to the left-hand side of Eq.~(\ref{phiF-nonlin-appr}), dividing the right-hand side, as each $0$-term is multiplied by a $k$-term in Eq.~(\ref{phiF-nonlin-appr}). Therefore, it should be defined via Eq.~(\ref{Fourier-nD0}) on the same step $s$ self-consistently
\begin{eqnarray}
\label{psi_0}
    \displaystyle \psi_0^{(s)} = \sqrt{1 - \sum\limits_{k\neq 0} \left[ \psi_{k}^{(s-1)} \right]^2} .
    %\displaystyle \psi_0^{(s)} = \sqrt{1 - \sum\limits_{k\neq 0} \left[ \tilde{\psi}_{k}^{(s-1)} \right]^2} .
\end{eqnarray}

\section{Calculation results and Discussion}
\begin{figure*}[htp]
(a)\includegraphics[scale=0.5]{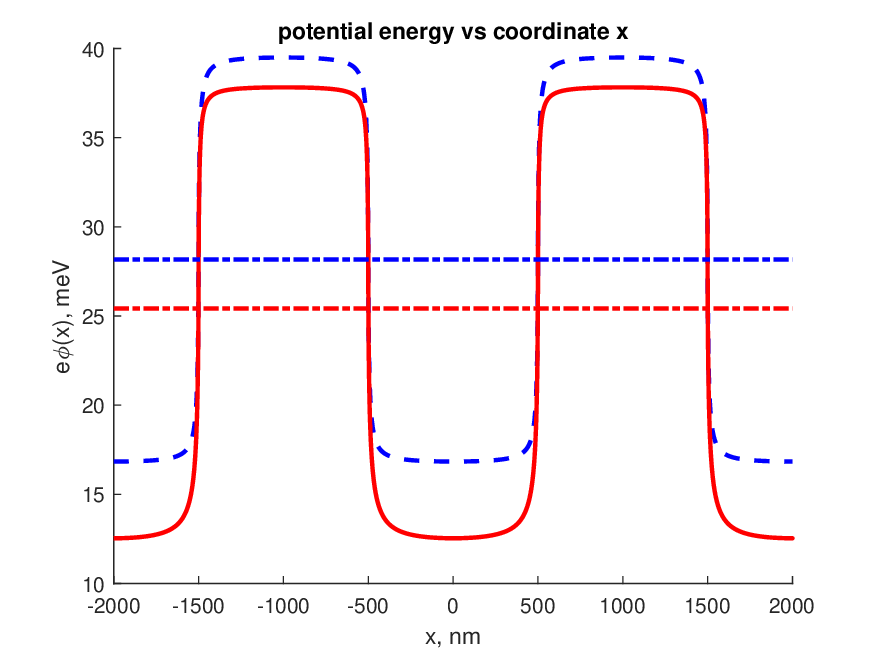}
(b)\includegraphics[scale=0.5]{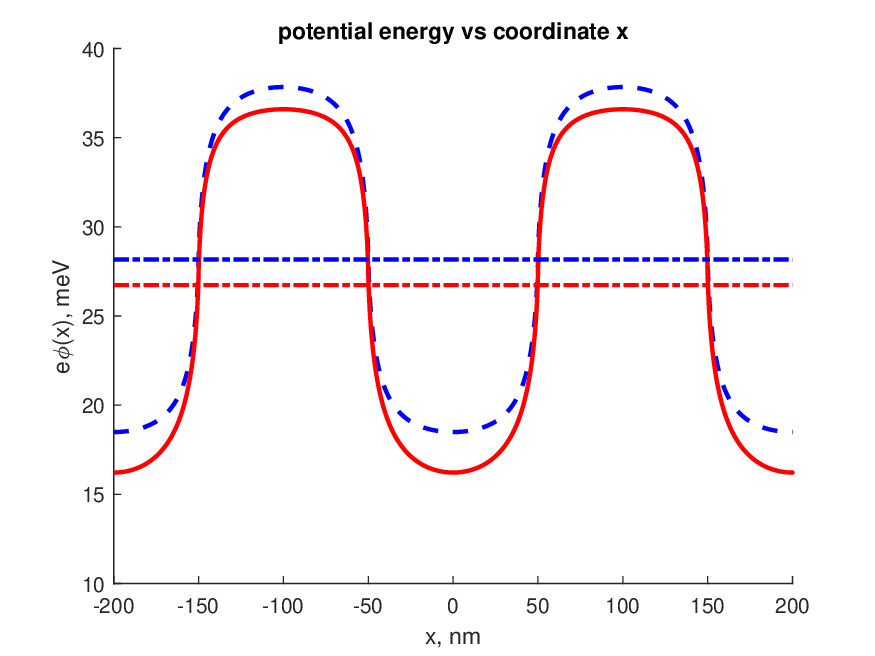}
\\
(c)\includegraphics[scale=0.5]{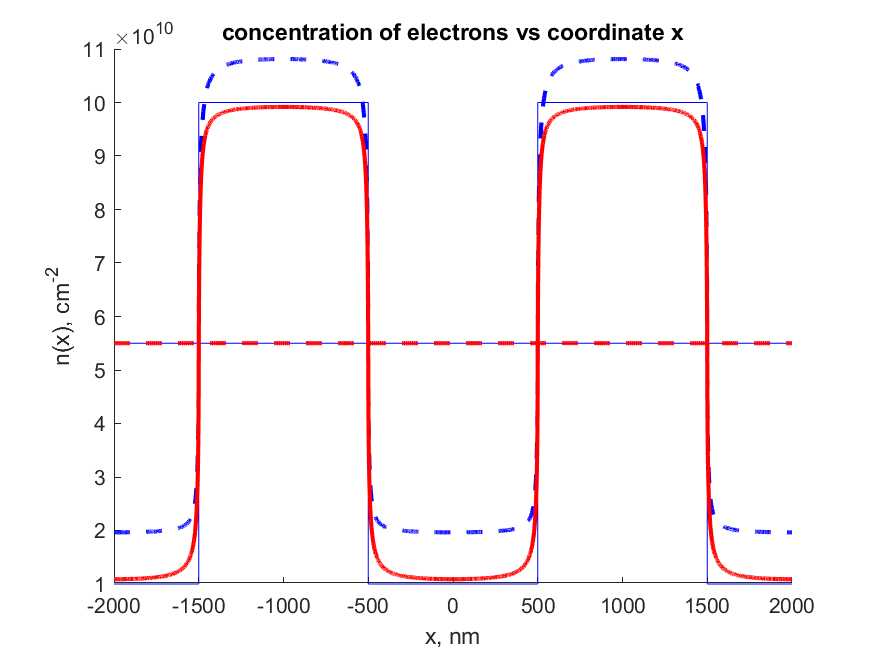}
(d)\includegraphics[scale=0.5]{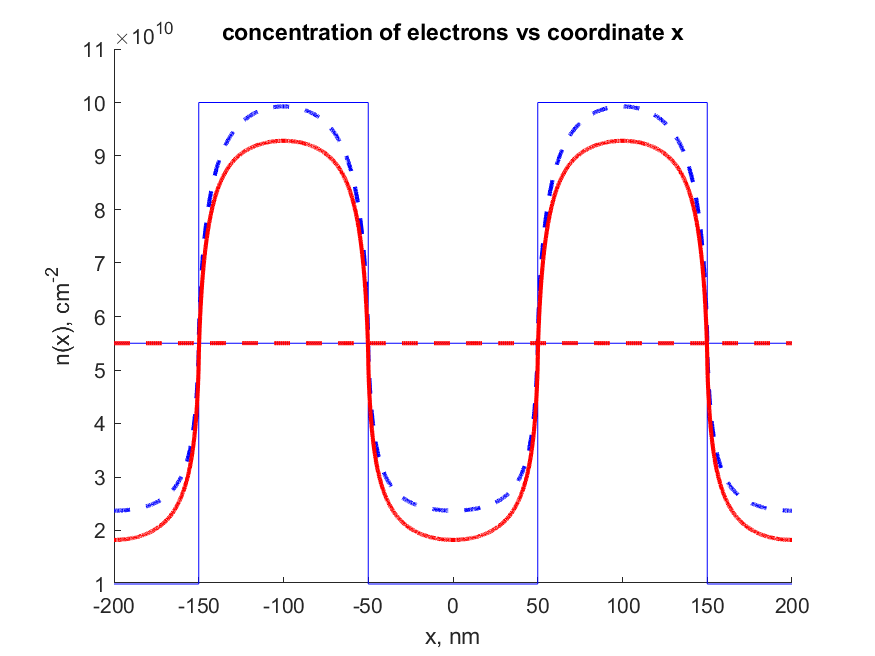}
\caption{\label{Res1} Absolute value of electrical potential energy $e\varphi(x)$ (panels a and b) and electron concentration $n(x)$ (panels c and d) along a periodically doped graphene sheet. The thick blue dashed and dash-dotted curves are calculated after the second iteration (equivalent to the linear approximation regime solution), while the thick red solid, dashed and dashed-dotted curves are calculated after $8$ iterations. The parameters used in the calculation are $n_{+} = 10^{11}$~cm$^{-2}$, $n_{0} = 10^{10}$~cm$^{-2}$, with panels a and c having $l= 1000 ~{\rm nm}$, $L=2000 ~{\rm nm}$, and panels b and d having $l= 100 ~{\rm nm}$, $L=200 ~{\rm nm}$, with $\varepsilon = 3$ in all cases. The dashed-dotted horizontal lines in panels a and b represent the mean values for potentials of corresponding colors. The horizontal lines in panels c and d represent the average concentration of donors (blue thin solid line) and the average concentration of electrons after $8$ iteration steps (dashed thick red line). The thin blue lines represent the change in donors concentration.}
\end{figure*}

The numerical scheme presented above provides a numerical solution within the framework of linear approximation equivalent to the analytical result in Eq.~(\ref{phiF-in-linappr}) at every second iteration step ($s=2$). The second iteration step is indicated by blue dashed curves in Figs.~\ref{Res1}(a--d). 
Therefore, when we consider a sample that satisfies the condition for the linear approximation regime $\Delta N \ll (n_{+} + n_{0})/2$, the solution does not change significantly with further iterations. For instance, if the concentrations of donors are such that $n_{0} = 9\cdot 10^{10} ~{\rm cm}^{-2}$ and $n_{+} = 10^{11} ~{\rm cm}^{-2}$, there is no difference between the second iteration step and the solution after $s=8$ iterations (all relative changes are less than $10^{-2}$ or all absolute changes for potential energy calculations are less than $10^{-2}$~meV, while the maximum change of potential energy along the sample sheet is $1.7$~meV; further iterations improve result by a relative change of $10^{-14}$). However, we are interested in cases with substantial differences in donor concentrations, where the linear approximation is not valid.

\begin{figure*}[htp]
(a)\includegraphics[scale=0.5]{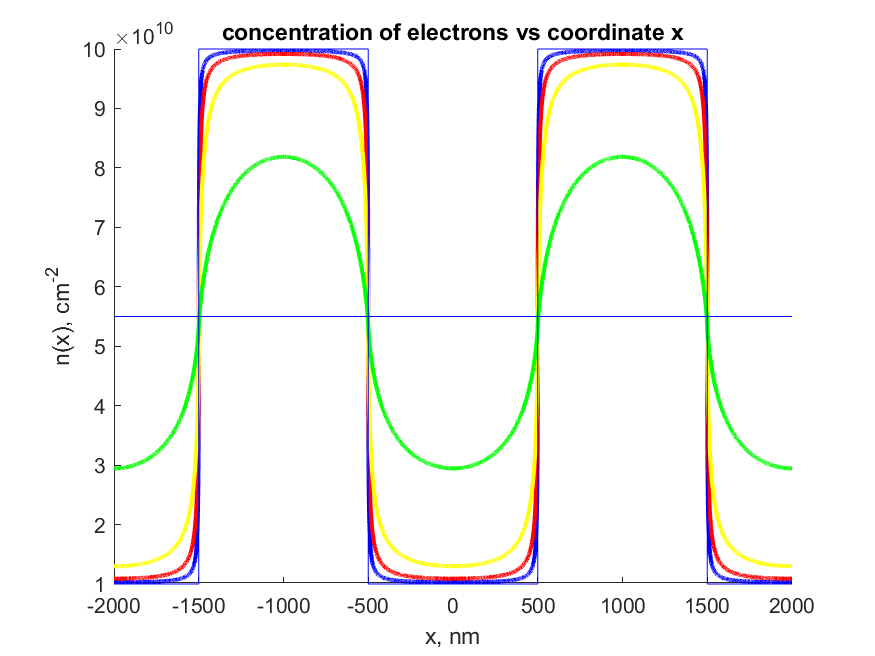}
(b)\includegraphics[scale=0.5]{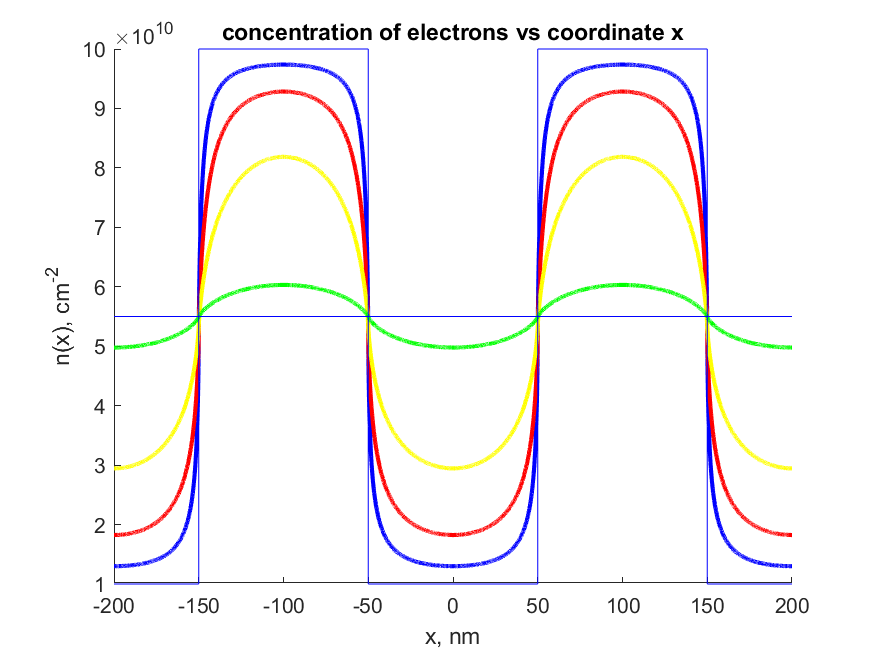}
\caption{\label{Res2} Electron concentrations along periodically doped graphene sheet. Thick solid curves are calculated after $8$ iterations. Parameters used in calculation are $n_{+} = 10^{11}$~cm$^{-2}$, $n_{0} = 10^{10}$~cm$^{-2}$, panel a: $l= 1000 ~{\rm nm}$, $L=2000 ~{\rm nm}$, panel b: $l= 100 ~{\rm nm}$, $L=200 ~{\rm nm}$. Different colors correspond to different values of outplane dielectric constants: blue -- $\varepsilon = 1$, red -- $\varepsilon = 3$, yellow -- $\varepsilon = 10$, green -- $\varepsilon = 100$. Thin blue lines represent donors concentration change along the sheet and also the mean value of donors concentration (horizontal line).}
\end{figure*}

Let us consider donor concentrations in the low-doped region as $n_{0} = 10^{10} ~{\rm cm}^{-2}$ and in the high-doped regions as $n_{+} = 10^{11} ~{\rm cm}^{-2}$ with equal widths of these regions $l/L=1/2$. The calculation results are presented in Figs.~\ref{Res1}--\ref{Res2}. In Fig.~\ref{Res1} one can see that the results differ at $s=2$ (blue dashed curves) and $s=8$ (red solid curves) with a relative change of values near $10^{-1}$ between these two cases, i.e. by tens of percents. Panels a and c in Fig.~\ref{Res1} represent calculation results for the dopants change period $L=2000$~nm, and panels b and d in Fig.~\ref{Res1} represent more narrow periods $L=200$~nm. Here we use $\varepsilon=3$ for the outplane dielectric constant. We stop the iteration process at $s=8$ as the relative changes for further iterations are less than $10^{-5}$. If one needs to improve the calculation precision, further iteration steps should be used (the process is convergent, while the requirement $\phi(x)>0$ holds at each step). 

One can see that the mean values of concentrations along the period for donors and electrons coincide with each other, revealing the validation of the electroneutrality condition at all iteration steps. Additionally, Fig.~\ref{Res1} illustrates that the average potential decreases with the growth of iteration number by several millielectronvolts. The change in potential energy itself along the period is $25$~meV in panel a and $20$~meV in panel b. This occurs becaus the electroneutrality condition in Eq.~(\ref{Fourier-nD0}) involves contributions from all Fourier components of the potential. As the nonzero Fourier components increase in absolute value during the iteration process, the zero Fourier component representing the average potential over a period decreases. This is because the electroneutrality condition acts as a normalization factor for contributions from all Fourier components. 

Fig.~\ref{Res2} shows the change in electron concentration along the periodically doped graphene sheet with the change of outplane materials with different dielectric constants $\varepsilon$. We consider only the symmetrical case here; the asymmetrical case with different materials on opposite sides of the graphene sheet plane does not drastically change the qualitative result (we can model this situation by substituting $\varepsilon$ with the algebraic average of dielectric constants of both sides). Panel a represents results for the $L=2000$~nm case, and panel b represents the $L=200$~nm case. Firstly, one can see that the narrower the period, the more prominent the inflow of electrons in low-doped regions. Secondly, the increase in $\varepsilon$ results in an increase in the electron concentration in low-doped regions and a decrease in high-doped regions. At $\varepsilon = 100$ (which lies between water $\varepsilon = 80$ at room temperatures \cite{artemov2019unified} and ice that can reach $\varepsilon = 190$ and greater values below $133$~K \cite{johari1975study,johari1981dielectric}), one can see a nearly homogeneous distribution of electrons in panel b. If one takes $\varepsilon = 1000$ (which is equivalent to bringing a metal closer to the graphene sheet) this distribution will differ from the mean value of donors by less than a percent, making it totally homogeneous. This illustrates an efficient electron redistribution due to the power-law of screening in the system. Such efficient inflow of electrons in low-doped regions can be used for fabricating high-mobility conducting channels.

\section{Conclusions}
The screened Coulomb potential of electrons in periodically doped graphene is significantly nonlinear. Under the assumption of weak spatial doping contrast, the problem is linearized and an analytical solution is derived. The precise solution of the problem involves the self-consistent determination of the Fermi level simultaneously with all Fourier components of the potential based on the electrical neutrality condition. A rapidly convergent iterative method for numerical calculation of significantly nonlinear problem is proposed. Calculations are performed for doping periods of $200$~nm and $2000$~nm with donor concentrations of $n_0 = 10^{10}$~cm$^{-2}$ and $n_+ = 10^{11}$~cm$^{-2}$. The potential energy difference reaches values of $20$ -- $25$~meV. By narrowing the doping period or bringing materials with high dielectric constants ($\varepsilon > 100$) in contact with the graphene sheet make, it becomes possible to create high-mobility channels in low-doped regions with a relatively high concentration of electrons redistributed from high-doped regions.

\section{Acknowledgements}
The work is done under financial support from Russian Science Foundation grant (Project 22-12-00211).

\providecommand{\noopsort}[1]{}\providecommand{\singleletter}[1]{#1}%

\end{document}